# MICROBURST APPLICATIONS OF BRIGHTNESS TEMPERATURE DIFFERENCE BETWEEN GOES IMAGER CHANNELS 3 AND 4

Kenneth L. Pryor
Center for Satellite Applications and Research (NOAA/NESDIS)
Camp Springs, MD

## 1. Introduction

This paper presents a new application of the brightness temperature difference (BTD) between Geostationary Operational Environmental Satellite (GOES) imager channels 3 and 4. It has been found recently that the BTD between GOES infrared channel 3 (water vapor, 6.5μm wavelength) and channel 4 (thermal infrared, 11μm wavelength) can highlight regions where severe outflow wind generation (i.e. downbursts, microbursts) is likely due to the channeling of dry mid-tropospheric air into the precipitation core of a deep, moist convective storm. Rabin et.al. (2010) noted that observations have shown that BTD > 0 can occur when water vapor exists above cloud tops in a stably stratified lower stratosphere. The physical process governing the effectiveness of the band 3-4 BTD in monitoring convective storm intensity is the forced ascent of water vapor to the lower stratosphere from overshooting tops. Thus, BTD > 0 has been used a measure for intensity of overshooting convection (Rabin et al. 2010). A new feature presented in this paper readily apparent in BTD imagery is a "dry-air notch" that signifies the channeling of dry air into the rear flank of a convective storm. A proven indicator of the presence of mid-level dry air and resultant downburst generation is the difference in equivalent potential temperature ($\Delta\theta_e$) between the surface and mid-troposphere (Atkins and Wakimoto 1991). Figure 1 graphically describes the evolution of the troposphere, as characterized by the $\theta_e$ profile, on a typical active microburst day in the southeastern United States. Case studies demonstrating effective operational use of this image product and associated parameters, such as $\Delta\theta_e$, are presented for two significant marine transportation accidents as well as a severe downburst event over the Washington, DC - Baltimore, Maryland metropolitan corridor in April 2010.

## 2. Methodology

The objective of this validation effort is to qualitatively and quantitatively assess the performance of the GOES imager channel 3-4 BTD in the indication of downburst activity by applying pattern recognition techniques. The image data consisted of derived brightness temperatures from infrared bands 3 and 4, obtained from the Comprehensive Large Array-data Stewardship System (CLASS, http://www.class.ncdc.noaa.gov/) Microburst algorithm output was visualized by McIDAS-V software (Available online at http://www.ssec.wisc.edu/mcidas/software/v/). A contrast stretch and built-in color enhancement ("Pressure") were applied to the output images to highlight patterns of interest including overshooting tops and dry-air notches. The dry-air notch identified in the BTD image is similar in concept to the rear-inflow notch (RIN) as documented in Przybylinski (1995). Visualizing algorithm output in McIDAS-V allowed for cursor interrogation of output brightness temperature and more precise recording of BTD values associated with observed downburst events.

For each downburst event, product images were compared to locations of marine transportation accidents, radar reflectivity imagery, and surface observations of convective wind gusts as provided by National Ocean Service observing stations (available via the National Data Buoy Center (NDBC)). Next Generation Radar (NEXRAD) base reflectivity imagery acquired from National Climatic Data Center (NCDC) was utilized to verify that observed wind gusts were associated with downbursts and not associated with other types of convective wind phenomena (i.e. gust fronts). Another application of the NEXRAD imagery was to infer microscale physical properties of downburst-producing convective storms. Particular radar reflectivity signatures, such as the bow echo and rear-inflow notch (RIN)(Przybylinski 1995), were effective indicators of the occurrence of downbursts.

Vertical profiles of $\theta_e$, relative humidity, and wind speed were derived from Global Forecast System (GFS) model output for the Concordia downburst event and from Rapid Update Cycle (RUC) model output for the April 2010 downburst event. In a similar manner to the BTD product images, the vertical profiles were visualized by McIDAS-V software. Features of interest to be identified in the vertical profiles included a large $\Delta\theta_e$ and associated mid-tropospheric $\theta_e$ minimum, a mid-level relative humidity minimum, and a wind speed maximum in the mid-tropospheric dry-air layer. All of these factors occurring simultaneously provided a high confidence that atmospheric conditions were favorable for downburst generation.

## 3. Case Studies

### 3.1 March 2004 Baltimore, Maryland Water Taxi Accident

During the afternoon of 6 March 2004, the "Lady D", a water taxi servicing the Baltimore Harbor, was capsized in a strong convective windstorm, resulting in five deaths. A cluster of convective storms developed over north-central Maryland during the afternoon, and then tracked rapidly eastward through the Baltimore Harbor and upper Chesapeake Bay between 2050 and 2120 UTC. Radar wind velocity measurements (not shown) from Dover AFB NEXRAD between 35 and 45 knots were associated with a cluster of downbursts as the convective storm complex approached the Baltimore Harbor. According to news reports, the Lady D capsized between 2050 and 2100 UTC.

GOES BTD imagery in Figure 2 clearly shows a dry-air notch on the western flank of a cluster of convective storms that was tracking from southern Carroll County and Howard County into western Baltimore County at 2032 UTC, about twenty minutes prior to the capsize of the Lady D. The dry-air notch corresponded very closely to the location of a RIN as identified in radar imagery at 2037 UTC. This suggests that the channeling of dry mid-tropospheric air into the rear flank of the storm and associated precipitation core was a major forcing factor for downdraft instability and the resulting downburst that occurred in Baltimore City and capsized the Lady D near 2050 UTC.

### 3.2 17 February 2010 S.V. Concordia Accident

During the afternoon of 17 February 2010, the Canadian Sailing Vessel (SV) Concordia sank off the coast of Brazil due to strong convective storm-generated winds. The SV Concordia was capsized in a downburst that occurred near 1722 UTC, about 290 nautical miles south-southeast of Rio de Janeiro (Capt. William Curry, SV Concordia, personal communication). GOES-12 Southern Hemisphere imagery was very effective in identifying the developing convective storm complex and favorable pre-conditions for downburst activity over one hour prior to the capsize of the Concordia. Satellite imagery indicated the presence of strong convective storm updrafts that resulted in heavy rainfall as well as the presence of a dry-air channel on the rear flank of the storm complex that most likely resulted in downburst generation.

Soden and Bretherton (1996) (SB96), in their study of the relationship of water vapor radiance and layer-average relative humidity, found a strong negative correlation between 6.5µm (channel 3) brightness temperature (BT) and layer-averaged relative humidity (RH) between the 200 and 500-mb levels. Thus, in the middle to upper troposphere, decreases in BT are associated with increases in RH as illustrated in Figure 4 of SB96. In the WV image in Figure 3, a notch of warmer brightness temperatures, indicated by the "V" pattern with orange shading on the southwestern flank of the storm complex, signified the presence of lower 500-mb humidity air being channeled into the rear of the storm.

The sequence of BTD images in Figure 4 between 1609 and 1709 UTC 17 February marks the location of the Concordia relative to the coast of Brazil. Also shown in the image sequence is the evolution of the thunderstorm complex that produced the strong downburst. The bright magenta shading in the thunderstorm complex indicates the presence of intense convection and associated strong updrafts that generated heavy rainfall. During this time period, a well-defined dry-air notch appears and tracks southeastward along the southwestern flank of the storm complex. This dry-air notch represents the drier (lower relative humidity) air that was channeled into the rear of the storm and provided the energy for intense downdrafts and the resulting possible downburst winds near 1720 UTC. The 3- and 6- hour forecasts of the 1200 UTC GFS model run illustrate the evolution of the atmosphere that promoted strong downdraft generation. The vertical profiles of $\theta_e$, relative humidity and wind speed at 1500 and 1800 UTC in Figure 5 show that over the 3-hour period surrounding the Concordia downburst, there was an expected increase in $\Delta\theta_e$, a marked decrease in mid-level relative humidity to below 20 percent, and an increase in wind speed in the mid-tropospheric dry air layer to near 25 knots. The trend in these parameters established favorable conditions for significant dry air entrainment into the developing thunderstorm complex. The 1$^{st}$ Mate of the Concordia reported a wind gust of 35 knots at 1721 UTC that was measured by a shipboard anemometer (Kimberley Smith, SV Concordia, personal communication). The magnitude of this wind gust is consistent with the winds associated with the capsize of the Lady D in the Baltimore Harbor.

Due to the absence of Doppler weather radar data for this case, only a tentative physical explanation of the downburst generation process can be provided. Entrainment of drier mid-tropospheric air into the precipitation core of the convective storm resulted in evaporation of precipitation, the subsequent cooling and generation of negative buoyancy (sinking air), and resultant acceleration of the downdraft. When this intense localized downdraft reached the ocean surface, downward vertical momentum was converted to horizontal momentum and resulted in strong outflow winds. The resulting strong winds then capsized the SV Concordia. Note that the dry-air notch was pointing directly to the location of the Concordia, and thus, the vessel was in the direct path of downburst winds. Also, it appears that the Concordia was in an optimal location to experience both the horizontal and vertical components of the downburst winds as illustrated in Figure 8. This suggests that the vessel experienced the early part (the contact to outburst stages) of a microburst.

**3.3 8 April 2010 Washington, DC/Maryland Bow Echo**

During the evening of 8 April 2010, a line of strong convective storms developed over northern Virginia and tracked northeastward over the Washington, DC and Baltimore, Maryland metropolitan areas between 2345 UTC 8 April and 0045 UTC 9 April 2010. The convective storm line evolved into a bow echo and produced widespread damaging winds in the Washington, DC area near 0000 UTC. GOES-12 imagery was very effective in identifying the evolution of the convective storm system and favorable pre-conditions for downburst activity at least fifteen minutes prior to the observation of severe winds in Washington, DC. Satellite and radar imagery indicated the presence of a dry-air channel on the rear flank of the bow echo (Przybylinski 1995) that most likely resulted in downburst generation.

The BTD images in Figure 6 mark the location of downbursts recorded at Washington, DC and Francis Scott Key (FSK) Bridge, Maryland National Ocean Service (NOS) observing stations. A well-defined southwest-to-northeast oriented dry-air notch appears on the southwestern flank of the bow echo that is co-located with a broad rear-inflow notch (RIN) as identified in radar imagery. This dry-air notch satellite feature, co-located with the RIN, most likely represents drier (lower relative humidity) air that was channeled into the rear of the storm and provided the energy for intense downdrafts and the resulting downburst winds at Washington (50 knots) near 0000 UTC and at FSK Bridge (42 knots) at 0036 UTC 9 April 2010. Note that the dry-air notch was pointing directly to the location of the observed downburst winds in both images. The line-end (or bookend) vortices apparent in radar imagery enhanced the channeling of dry-air into the rear flank of the convective storm. By the time the storm was tracking over the Baltimore area, the dry-air notch appeared to be embedded within the eastern line-end vortex, which is a region in the storm that is favorable for downburst generation.

Comparison of product images to transmittance weighting functions and a radiosonde observation (RAOB) at Dulles International Airport, Virginia at 0000 UTC 9 April emphasizes the importance of the dry air in the rear inflow region in the generation of intense downdrafts. Figure 7 shows a steep, near dry-adiabatic temperature lapse rate in the mid-troposphere, between the 300 and 500-mb levels, and a co-located dry-air layer with a mixing ratio near zero. The channel 3 weighting function peak near the 400-mb level and channel 4 weighting function peak near the surface is associated with an overall steep temperature lapse rate between the 300-mb level and the surface. More information pertaining to transmittance weighting functions is available at this site. Thus, the maximum BTD value near 30K observed at 0015 UTC near Dulles Airport can be related to a convectively unstable environment with sufficient mid-level dry air to generate strong downdrafts as the dry air is channeled into the rear flank of the convective storm line and heavy precipitation core. A correlation has been found between maximum BTD in the dry air notch and corresponding downburst wind gust magnitude: 39K over northern Virginia at 2345 UTC associated with the 50-knot gust observed at Washington, DC at 0000 UTC; and 30K southwest of Baltimore associated with the 42-knot gust observed at FSK Bridge at 0036 UTC. This entrained dry air results in evaporation of descending precipitation and cooling within the downdraft, and the subsequent acceleration of the downdraft toward the surface. Similar to the Concordia downburst case, the vertical profiles of $\theta_e$ and relative humidity displayed favorable atmospheric conditions for downburst generation that included a well-defined dry-air layer with strong winds of 30 to 40 knots that enhanced dry air entrainment. A graphical description of the physical process of downburst generation within a bow echo is presented in Figure 8.

Downward horizontal momentum transport may have also been another factor in downburst generation and magnitude. The wind profile in the Dulles Airport RAOB at 0000 UTC 9 April (not shown) indicated wind direction from the southwest (230°) and a wind speed near 50 knots in the downdraft entrainment region near the 400-mb level. The orientation of the dry-air notch (240°) corresponded even more closely to the wind direction (260°) associated with the Washington, DC downburst.  Similarly, the wind direction (250°) associated with the downburst wind gust at FSK Bridge corresponded closely with the orientation (230°) of the dry-air notch between Washington, DC and Baltimore.  The 20° deviation between dry-air notch orientation and downburst wind gust direction at the surface is likely the result of the divergent nature of the convective storm outflow on the surface.  The correlation between the direction of orientation of the dry-air notch and the direction of observed downburst winds on the surface emphasizes the importance of the dry-air notch in the enhancement of convective storm downdrafts and resulting storm outflow.

 4. Discussion and Conclusion

The dry-air notch identified in all three cases presented above most likely represents drier (lower relative humidity) air that was channeled into the rear of convective storms and associated precipitation cores, and subsequently provided the energy for intense downdrafts and resulting downburst winds. In two cases, comparison of BTD product imagery to corresponding radar imagery revealed a correlation between the dry-air notch and the RIN.  Entrainment of drier mid-tropospheric air into the precipitation core of the convective storm typically results in evaporation of precipitation, the subsequent cooling and generation of negative buoyancy (sinking air), and resultant acceleration of a downdraft. When the intense localized downdraft reaches the surface, air flows outward as a downburst. Ellrod (1989) noted the importance of low mid-tropospheric (500 mb) relative humidity air in the generation of the severe Dallas-Fort Worth, Texas microburst in August 1985.  It is noteworthy that BTD product imagery has the ability to infer the immanence of downbursts in the absence of radar reflectivity, especially over open ocean waters as was demonstrated in the Concordia downburst case.

5.  References

Atkins, N.T., and R.M. Wakimoto, 1991: Wet microburst activity over the southeastern United States: Implications for forecasting. *Wea. Forecasting*, **6**, 470-482.

COMET, 1999: Mesoscale Convective Systems: Squall Lines and Bow Echoes. Online training module, http://www.meted.ucar.edu/convectn/mcs/index.htm.

Ellrod, G. P., 1989: Environmental conditions associated with the Dallas microburst storm determined from satellite soundings. *Wea. Forecasting*, **4**, 469-484.

Przybylinski, R.W., 1995: The bow echo. Observations, numerical simulations, and severe weather detection methods. *Wea. Forecasting*, **10**, 203-218.

Rabin, R., P. Bothwell, and S. Weiss, 2010: Temperature deviation from Equilibrium Temperature: Convective Overshoot from Satellite Imagery. [Available online at http://overshoot.nssl.noaa.gov/.]

Soden, B.J. and F.P. Bretherton, 1996: Interpretation of TOVS water vapor radiances in terms of layer-average relative humidities: Method and climatology for the upper, middle, and lower troposphere. *J. Geophys. Res.*, **101**, 9333-9343.

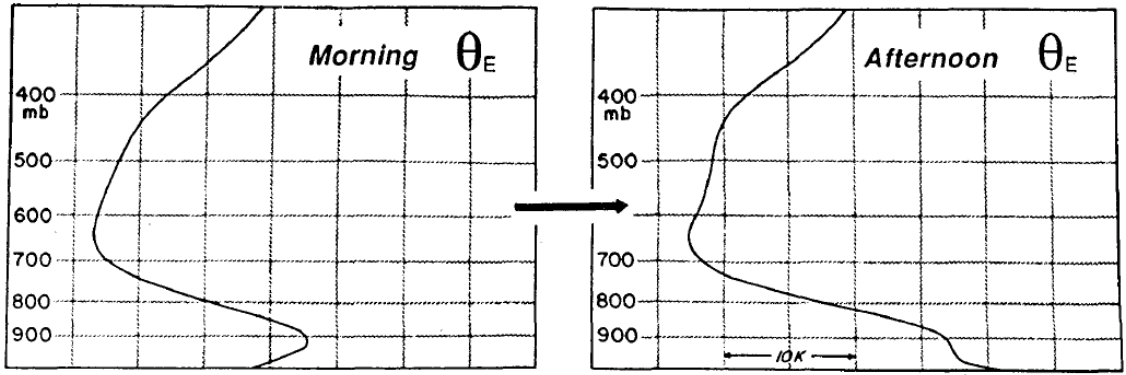

Figure 1. Schematic $\theta_e$ profiles on days when the environment is conducive for wet microburst occurrence in a humid region (after Atkins and Wakimoto 1991).

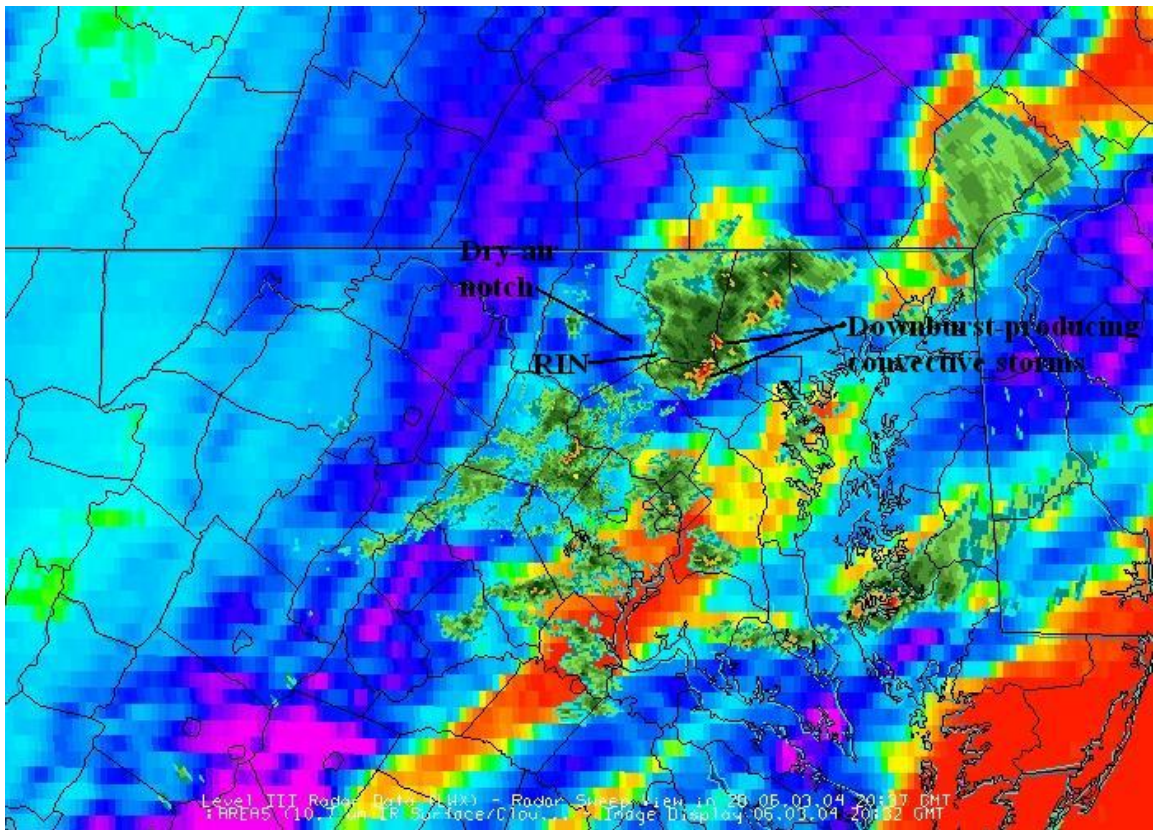

Figure 2. GOES BTD product image at 2032 UTC 6 March 2004 with overlying radar reflectivity image from Sterling, Virginia NEXRAD at 2037 UTC. Location of the Baltimore Harbor is marked with an "X".

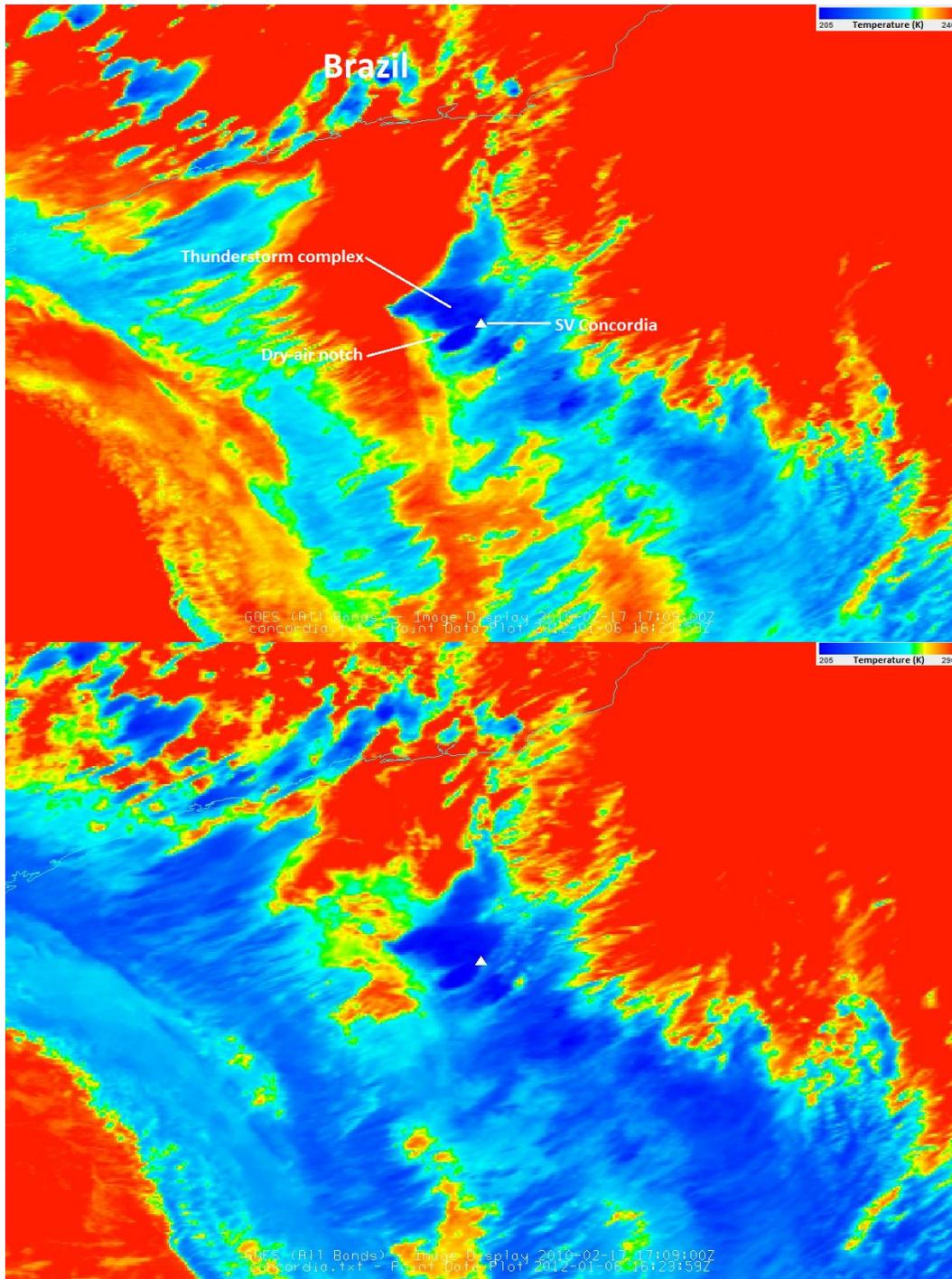

Figure 3. GOES-12 water vapor (WV, top) and thermal infrared (IR, bottom) imagery at 1709 UTC 17 February 2010. Location of the SV Concordia is indicated by a white triangle.

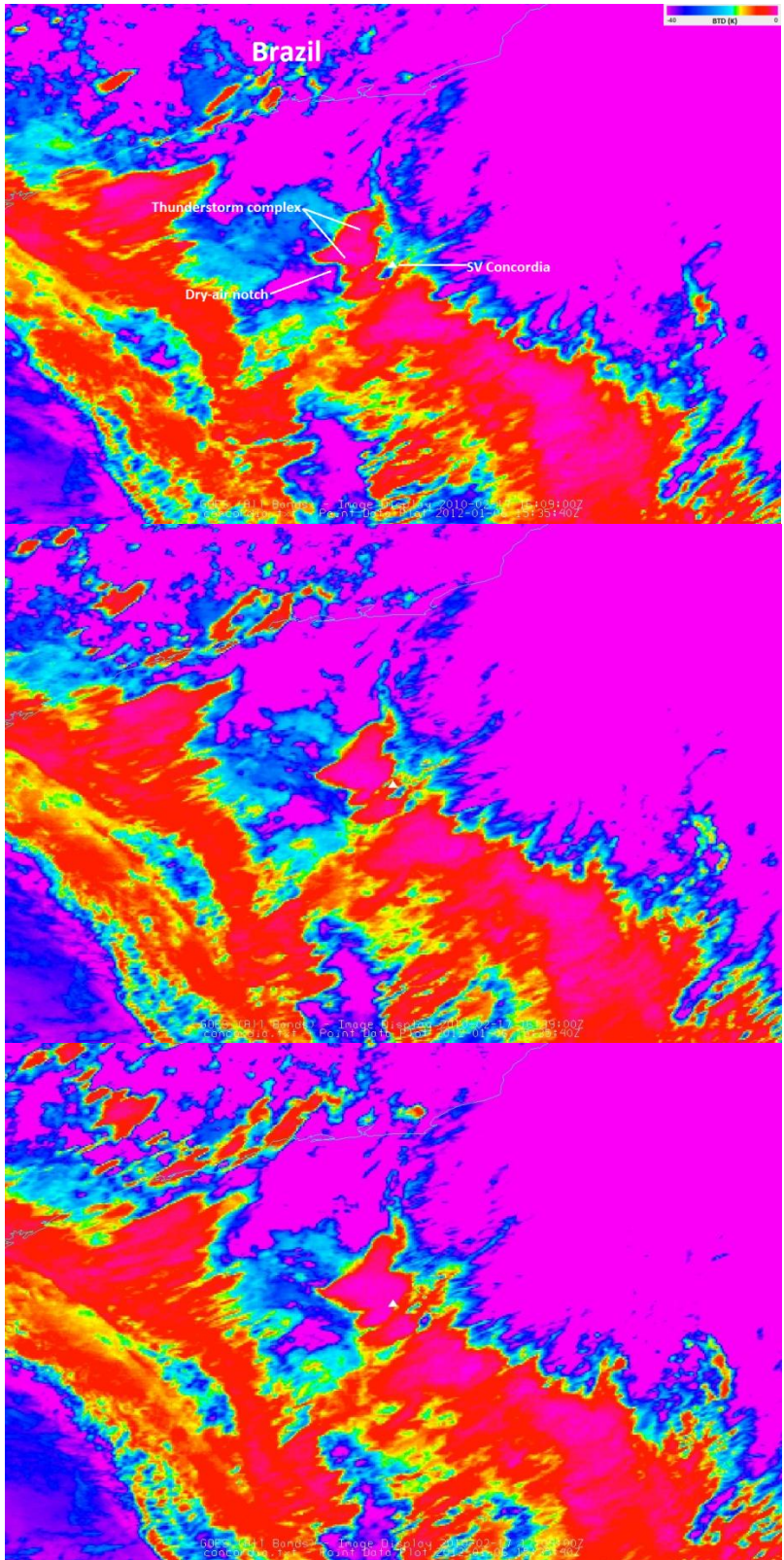

Figure 4. GOES-12 channel 3 (WV)-channel 4 (IR) brightness temperature difference (BTD) product images at 1609 UTC (top), 1639 UTC (middle), and 1709 (bottom) 17 February 2010.

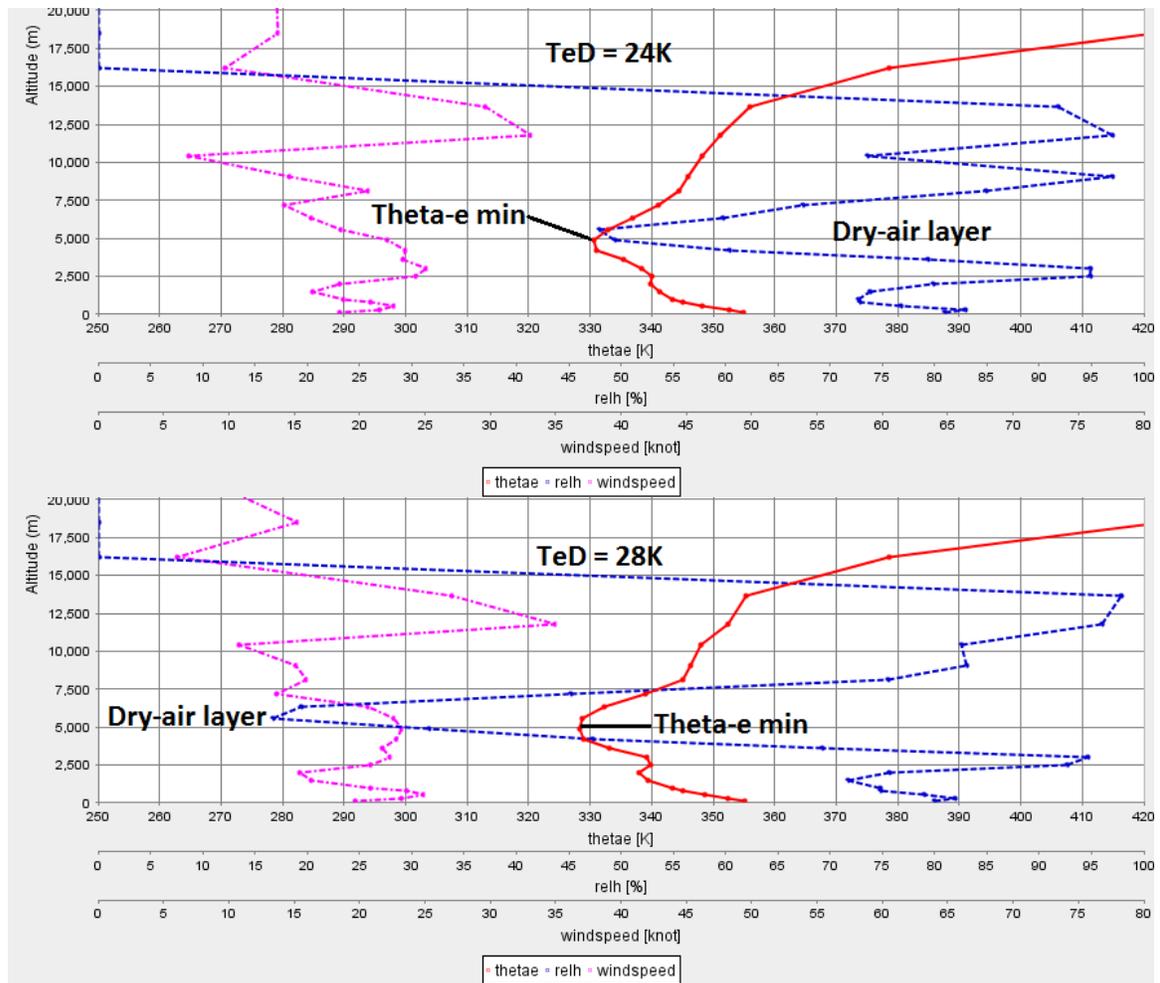

Figure 5.  Vertical profiles of relative humidity (blue), theta-e (red), and wind speed (knots, pink) over the location of the Concordia downburst at 1500 UTC (top) and 1800 UTC (bottom) 17 February 2010.

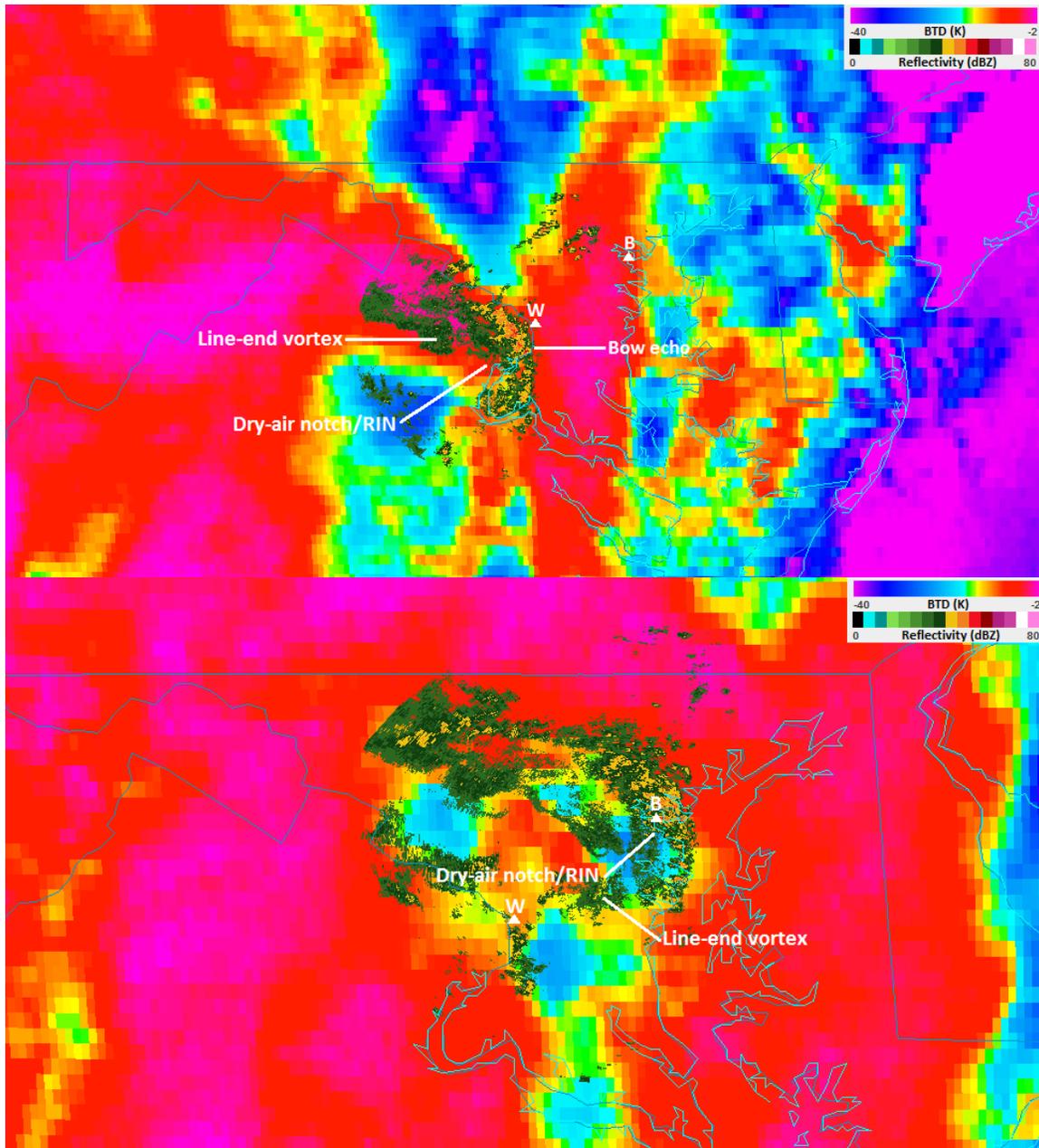

Figure 6. GOES-12 channel 3 (WV)-channel 4 (IR) brightness temperature difference (BTD) products at 2332 UTC 8 April (top) and 0102 UTC 9 April 2010 (bottom) with overlying radar reflectivity from Washington Terminal Doppler Weather Radar (TDWR) at 2349 UTC (top) and Baltimore-Washington International Airport TDWR at 0040 UTC (bottom). White line represents the co-located dry-air notch and rear-inflow notch (RIN) pointing to Washington, DC National Ocean Service (NOS) observing station (location "W") and Francis Scott Key Bridge, Maryland NOS observing station (location "B").

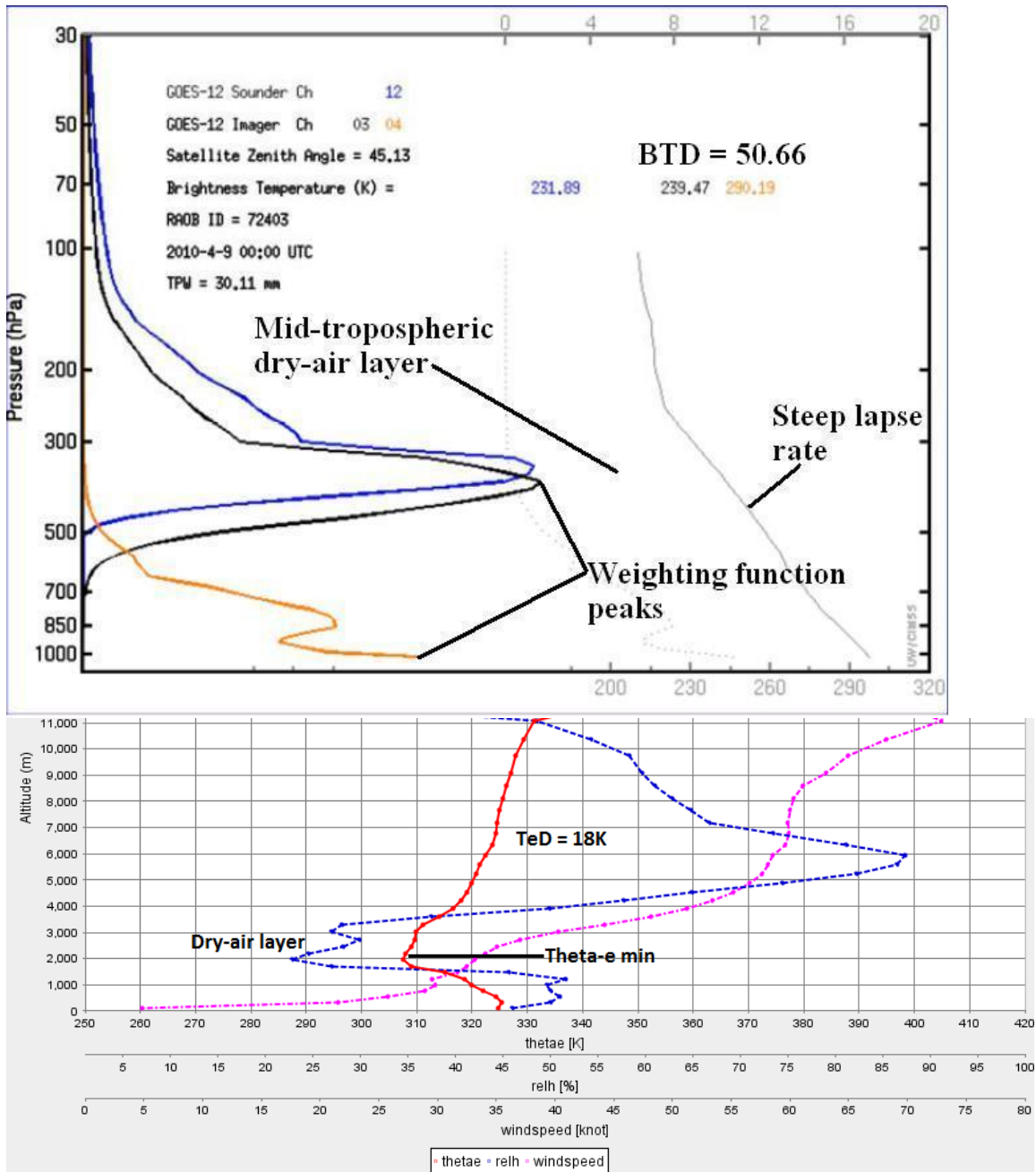

Figure 7. Plot of transmittance weighting functions for channels 3 and 4 and temperature and mixing ratio curves from the 0000 UTC RAOB at Dulles Airport (top) compared to a vertical profile of relative humidity (blue), theta-e (red), and wind speed (knots, pink) over Washington, DC at 2300 UTC 8 April.

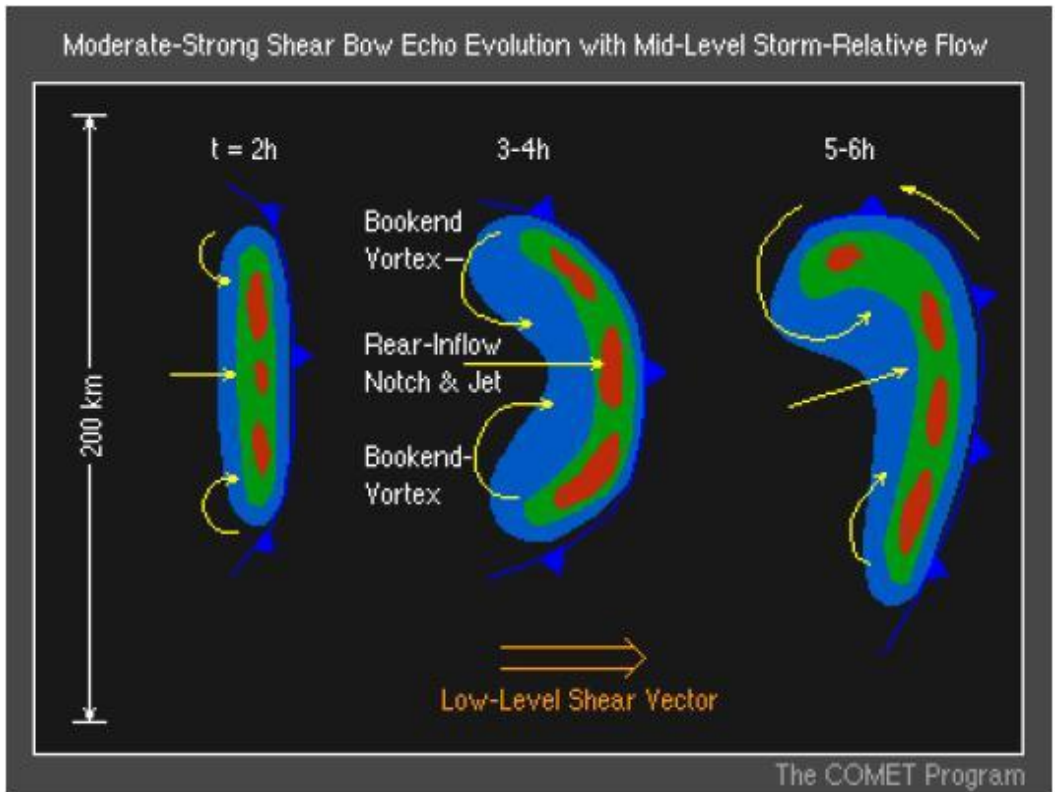
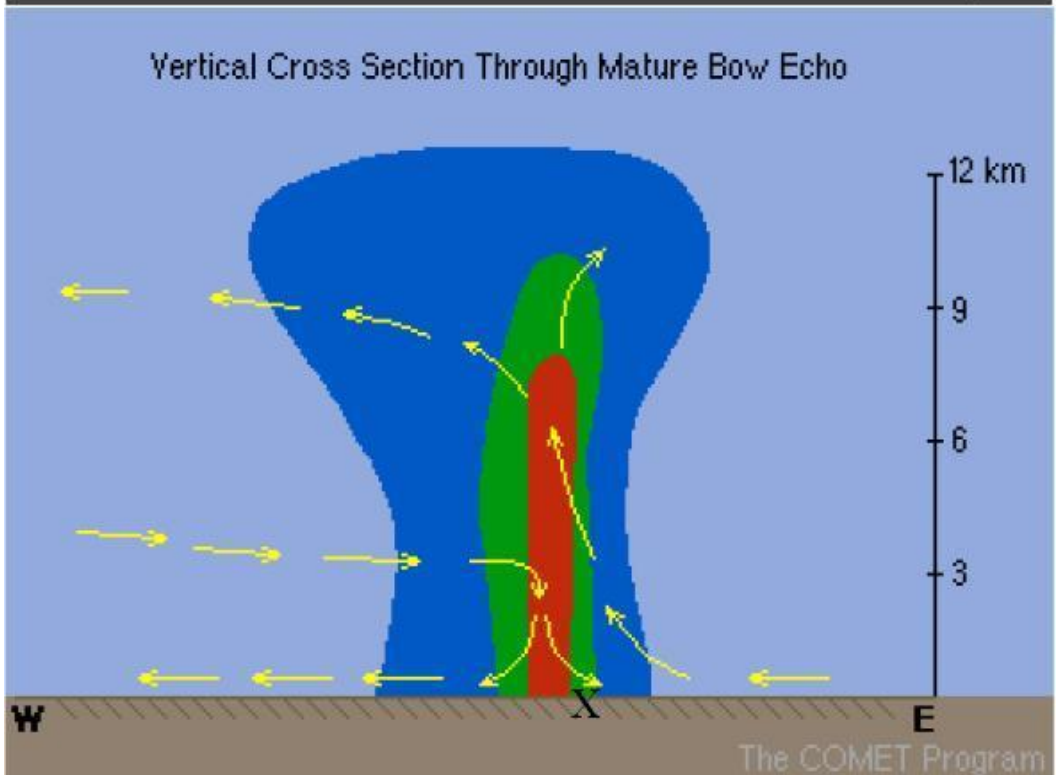

Figure 8. Conceptual model of a strong bow echo evolution showing bookend vortices and development of a Rear-Inflow Notch (RIN) (top) and schematic of a vertical cross-section through a mature bow echo (bottom). Location "X" marks the likely position of SV Concordia. Courtesy of COMET (1999).